\documentclass[reprint,superscriptaddress,
 amsmath,amssymb, aps, 
 usenames,dvipsnames]
 {revtex4-1}
 
 \usepackage[dvipsnames]{xcolor}
 \usepackage{xr}

 \usepackage[titletoc,toc,title]{appendix}

\usepackage{hyperref}
\usepackage{graphicx,epstopdf,bm,amssymb,amsfonts,dcolumn,bm,xcolor,xr}
\hypersetup{colorlinks=true,linkcolor=blue,citecolor=red}

\begin{document}		

\title{Diversity Improves Speed and Accuracy in Social Networks}
\author{Bhargav Karamched}
\email{bhargav@math.uh.edu}
\affiliation{Department of Mathematics, University of Houston, Houston, Texas 77004, USA}
\author{Megan Stickler}
\affiliation{Department of Mathematics, University of Houston, Houston, Texas 77004, USA}
\author{William Ott}
\affiliation{Department of Mathematics, University of Houston, Houston, Texas 77004, USA}
\author{Benjamin Lindner}
\affiliation{Physics Department, Humboldt University, Berlin, Germany}
\affiliation{Bernstein Center for Computational Neuroscience, Berlin, Germany}
\author{Zachary P. Kilpatrick}
\affiliation{Department of Applied Mathematics, University of Colorado Boulder, Boulder, Colorado 80309, USA}
\author{Kre\v{s}imir Josi\'{c}}
\affiliation{Department of Mathematics, University of Houston, Houston, Texas 77004, USA}
\affiliation{Department of Biology and Biochemistry, University of Houston, Houston, Texas 77004, USA}
\email{josic@math.uh.edu}



%

\date{\today}


\begin{abstract}
How does temporally structured private and social information shape collective decisions?  To address this question we consider a network of rational agents who independently accumulate private evidence that triggers a decision upon reaching a threshold. When seen by the whole network, the first agent's choice initiates a wave of new decisions; later decisions have less impact.   In heterogeneous networks, first decisions are made quickly by impulsive individuals who need little evidence to make a choice, but, even when wrong, can reveal the correct options to nearly everyone else.  
We conclude that groups comprised of diverse individuals can make more efficient decisions than homogenous ones.
\end{abstract}

\pacs{Valid PACS appear here}

\maketitle


A central question in biology, sociology, and economics is how the exchange of information shapes group decisions~\cite{Couzin2009, Edwards54, Frith2012,PE2011,Arganda12,Mann18,Mann2020}.  
Humans and other animals observe the choices of their peers to guide their own decisions~\cite{Ward08, Ward12, Gall17,SP15}:  Argentinian ants form trails by following their peers \cite{Perna12}, African wild dogs depart a congregation in response to their neighbor's sneezes~\cite{Walker17},
and pedestrians look to each other to decide when to cross a road~\cite{Faria10}.

How do individuals combine private evidence and social information to make decisions?  What should they do when they observe choices at odds with their own beliefs?   To address these questions we propose an analytically tractable model of collective, rational decision-making. 
Agents in the network accumulate private evidence according to the widely-adopted drift-diffusion model (DDM) \cite{Ratcliff1978theory,Gold02,Bogacz2006}. They do not share private information, but observe each other's choices~\cite{Mann18,Karamched20}.   A decision reveals the evidence an agent accumulated and may trigger decisions by undecided observers.

We show that in a group of identical agents, a wrong first decision leads approximately half the network astray.  However, in heterogeneous networks a wrong first choice is usually made by a hasty, uninformed agent and only convinces others who are similarly quick to decide. More cautious agents can observe the decisions of these early adopters, and make the right choice.
We conclude that in diverse groups  decisions by unreliable agents, even when wrong, can reveal the better option.

\begin{figure}
  \includegraphics[width=1.01\columnwidth]{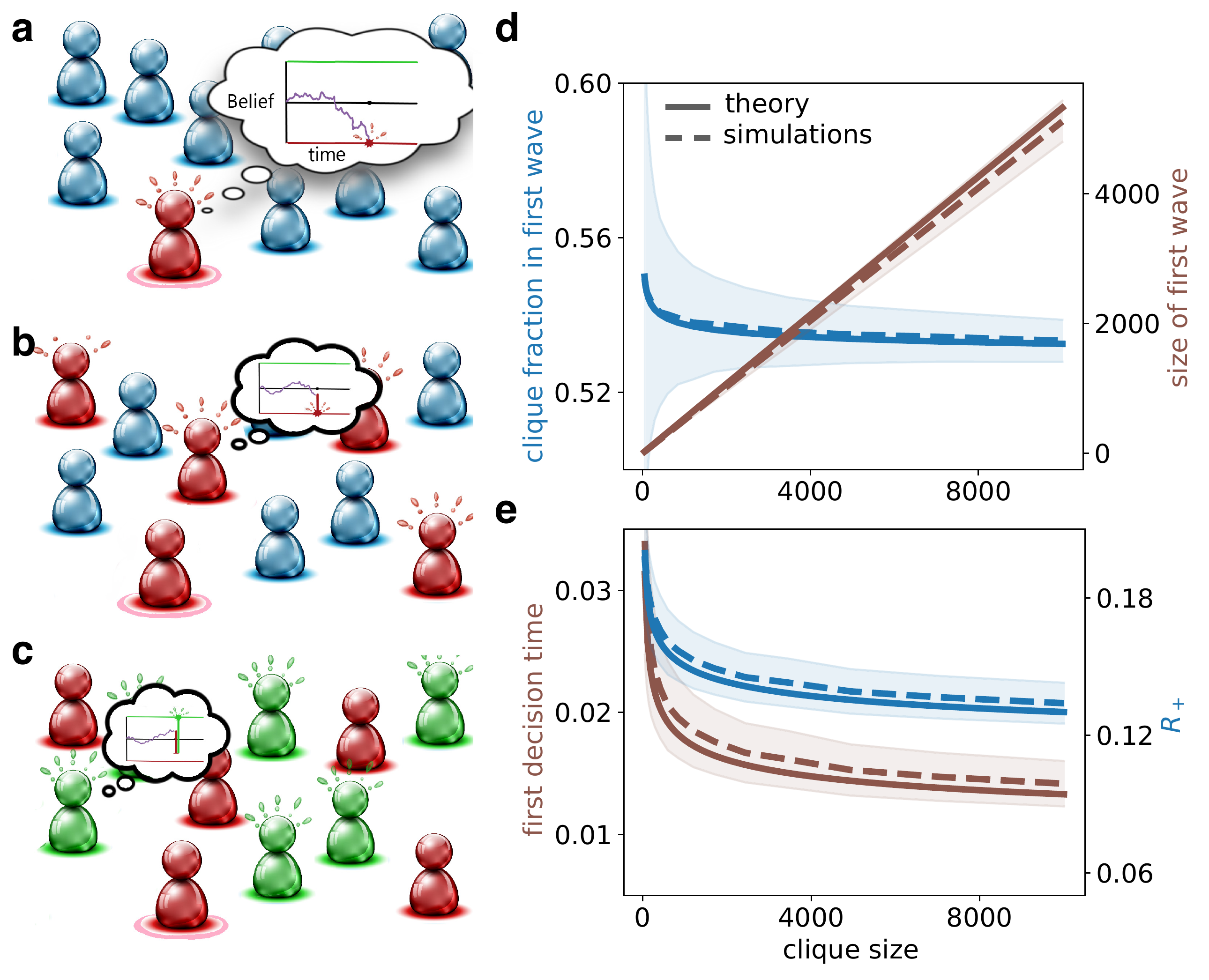}
  \caption{(Color online) {\bf Waves of collective decisions.}  (a)~The first in a clique of identical agents gathers sufficient private evidence but decides incorrectly~(red).  (b)~The first decision convinces a small wave of agents to agree.  Since this wave is small, it reveals to the undecided~(blue) agents that the first decision was likely wrong. (c)~The difference between decided and undecided agents leads a second wave of agents to choose correctly (green).  (d) The first wave increases with network size (red), but comprises a smaller fraction of the population (blue; See Eq.~\eqref{a11}).  (e) The time to the first decision decreases with network size (red), allowing each agent less time to accumulate private information (See Eq.~\eqref{first_decision1}).  The amount of information provided by a first wave decision  also decreases ($R_+$, blue; See Eq.~\eqref{E:update}).  We used $\theta = 0.7$ in (d) and (e).  Here, and below, solid and dashed lines represent simulations and theory, respectively, and shaded regions capture one standard deviation of simulated results around the mean. 
  }
  \vspace{-6mm}
 \label{Fig:1}
\end{figure}

\emph{Model Description.} We consider an all-to-all network, or {\em{clique}}, of agents, each deciding between two options (Fig.~\ref{Fig:1}). To do so, agents continuously accumulate private evidence and social evidence from other agents.  The agents do not share their private observations, but know the statistics of the observations each agent makes, and  
observe the choices of all agents in the network.  A decision, once made, is final and cannot be undone.

Before giving precise definitions, an example provides some intuition:  A group of people is deciding which one of two 
new products to buy.  To do so, they study the products' specifications, examine their performance, and
read reviews, making a sequence of observations of varying reliability. 
They also observe which product their friends choose to buy.  To decide, each person combines their private observations (product reviews) with social information (purchase decisions of friends). They do not exchange information before making a purchase, but know the type of information their friends gather, and thus the statistics of how beliefs evolve~\cite{Goyal2012,Mann18}. 
Once a purchase is made, the product cannot be returned.

\paragraph{Evolution of observers' beliefs.} 
We assume $N$ agents (observers) accumulate noisy  private observations and optimally combine them with  information obtained from observing the decisions of their neighbors  to choose between two hypotheses, $H^+$ or $H^-$.  Either hypothesis is \emph{a priori} equally likely to be correct.  Each agent, $i,$ makes decisions based on their \emph{belief}, $y_i(t),$ which equals the log likelihood ratio (LLR) between the hypotheses given all available evidence.  After making a sequence of private 
observations,  $\xi^{(i)}_{1:t},$ the belief is therefore $y_i(t)= \log [ \mathbb{P} (H^+ | \xi^{(i)}_{1:t}) /  \mathbb{P} (H^- | \xi^{(i)}_{1:t})]$.  If private observations are rapid and \emph{uncorrelated} in time and between agents, each agent's belief approximately evolves as 
\begin{equation}
dy_{i} = \pm \alpha dt + \sqrt{2 \alpha} dW_{i},
\label{langevin}
\end{equation}
where the sign of the drift equals that of the correct hypotheses and $W_i(t)$  are independent, standard Wiener processes~\cite{Bogacz2006, veliz16}. 
 Each observer starts with no evidence or bias, so $y_{i}(0) = 0$.  We assume henceforth that $H^+$ is correct, and that the drift in Eq.~\eqref{langevin} is $\alpha = 1$. When $H^-$ is correct and $\alpha \neq 1$ the analysis is similar. 

%

Each agent, $i,$ sets a threshold, $\theta_i$, and chooses $H^+$ ($H^-$) at  time $T_i$  if $y_{i}(T_i) \geq  \theta_i \; \; (y_{i}(T_i) \leq  -\theta_i)$, and $y_{i}(t) \in (-\theta_i, \theta_i)$ for $ 0 \leq t < T_i$.    An agent's decision is observed by all other agents and cannot be undone. 

An agent that observes a decision will know whether the decider chose $H^+$ or $H^-,$ but may not know the threshold of the decider.  We will  consider {\em{omniscient}} agents who know each other's thresholds, and the case of {\em{consensus bias}} where agents assume all others have the same thresholds they do.



\paragraph{Belief updates due to a decision.} Without loss of generality, we assume the belief of agent $i = 1$ is the first to reach threshold at time  $t = T$ (See Fig.~\ref{Fig:1}a). The probability that this decision is correct is $(1 + \exp(-\theta))^{-1}$~\cite{Bogacz2006}.  

Until the first decision, beliefs of all agents, $y_{i}(t),$ evolve independently according to Eq.~\eqref{langevin}. Upon observing the first decision, \emph{omniscient} agents update their belief by the amount of evidence independently accumulated by the first decider,   $y_i(T) \rightarrow  y_i(T) \pm \theta_1$ \cite{Karamched20}. A positive ($H^+$) first decision ($y_1(T) = \theta_1$) and update causes any belief that satisfies $y_{i}(T^-) \in [\theta_i - \theta_1, \theta_i)$ just before the first choice,  to cross the positive threshold, $\theta_i$, evoking a positive decision by agent $i$. Similarly, agents whose belief satisfies $y_{i}(T^-) \in (-\theta_i ,  \theta_1 - \theta_i]$ would follow a negative first decision.  Agents subject to \emph{consensus bias} update their belief as
$y_i(T) \rightarrow  y_i(T) \pm \theta_i$.
For simplicity we assume  agents exchange all social information before accumulating further private information.
%

Multiple waves of decisions can now follow: The first choice is followed by a wave of $a_1$ \emph{agreeing} agents (See Fig.~\ref{Fig:1}b). Each of the $N - a_1 - 1$ undecided agents obtains information by observing who followed the first decision and who did not. How do the remaining rational agents make use of this newly revealed information?

\emph{Homogeneous populations.}
To answer this question, first suppose all agents have identical thresholds, $\theta_i=\theta$, for all $i$, so that the cases of omniscience and consensus bias are equivalent. 
Observing that agent $i \neq 1$ follows a positive first decision tells other agents that $y_{i}(T^-) \in [0, \theta)$. An agent's belief equals the LLR of the conditional probabilities of the two options, given all available information. Therefore  observing first wave decision of agent $i$   leads to an increment in belief  equal to~\cite{Karamched20}
\begin{align*}
\text{LLR}(y_{i}(T) \in (0, \theta)) &\overset{\mathrm{def}}{=} \log\left(\frac{\mathbb{P}(y_{i}(T) \in [0, \theta) | H^+) }{\mathbb{P}(y_{i}(T) \in [0, \theta) | H^-)} \right) \\
&=\log\left(\frac{\int_{0}^{\theta}p_+(x,T) dx}{\int_{0}^{\theta}p_-(x,T) dx}\right) \equiv R_+(T).
\end{align*}
Here $p_{\pm}(x,t) \Delta x =  P(y_i(t) \in (x, x + \Delta x) | H^{\pm}) + o( \Delta x)$ is the conditional probability density for the belief of agent $i$ 
at time $t$. Since thresholds are symmetric, $\int_{0}^{\theta} p_-(x,t) dx = \int_{-\theta}^0 p_+(x,t) dx$, so observing an agent $j$ who remains \emph{undecided} after  the first decision reveals $y_{j}(T) \in (-\theta, 0]$, leading to an 
increment  
$
\text{LLR}(y_{j}(T) \in (-\theta, 0]) \equiv R_-(T) = -R_+(T)$.

We assume that agents know the statistics of private observations and can therefore compute $p_{+}(x,t)$ and $p_{-}(x,t)$.  
Agents thus know that beliefs evolve according to Eq.~\eqref{langevin} and that the belief distribution prior to any decision satisfies:
\begin{align}
\label{smoluchowski}
\partial_t p_{\pm}  =  \mp \partial_x p_{\pm}  + \partial^2_{xx} p_{\pm}, \qquad
p_{\pm}(\pm \theta, t) = 0, 
\end{align} 
if $H^{\pm}$ is correct, with $p_{\pm}(x,0) = \delta(x)$. Agents do not know which hypothesis is correct, and compute the belief update, $R_+(T),$ using the belief distributions, $p_{\pm}(x,t)$.

If the first agent chooses $H^+$,  agents undecided after the first wave  combine the information from all observed decisions and indecisions.
Since private measurements are independent, information obtained from agents in the first wave is additive,  and the 
belief increment is 
\begin{align}
c_1^+ &\overset{\mathrm{def}}{=} a_1 R_+(T) + (N - a_1 - 2)R_-(T)\nonumber \\
&= \left(2 a_1 - N  + 2 \right)R_+(T).
\label{E:c1}
\end{align}

If the first decision sways more than half the network, $a_1 > N/2 -1,$ the weight of  new evidence favors the choice of the first agent. Conversely, observing that the majority of agents remain undecided provides evidence against the first agent's decision. 
All undecided agents increment their belief by  $ c_1^+,$ leading to a second wave of $a_2$ decisions, all of equal sign (See Fig.~\ref{Fig:1}c). Agents in the first wave agree with the first decision, while agents in the second wave agree only if the sign of $c_1^+$ matches the first decision.  Observers still undecided after this second wave update their LLR by a new increment $c_2^+$, and waves of decisions follow until either all agents make a choice, or  no new agent makes a decision after some belief update, $c_k^+$, $k \geq 2$. Undecided agents then continue to accumulate private information. Whether the first decision is right or wrong, we will show that in large populations the second wave of decisions encompasses the entire population. 

If the first agent wrongly chooses  $H^-$, the computations are  similar:  Observing a  decision in the first wave provides a belief increment $R_-(t) = - R_+(t)$,  and observing an undecided agent provides an increment  $R_+(t),$ giving a total increment 
$c_1^-  = \left(2 a_1 - N  + 2 \right)R_-(T)$.  Further decision waves follow equivalently. 

\emph{a. Decisions in large groups.} 
As $N$ grows, the time to the first decision, $T,$ approaches 0, and we can approximate the solution to Eq.~(\ref{smoluchowski}) using the method of images~\cite{cox65,Drugowitsch16} . Using the resulting lifetime distribution and extreme value theory we find that the expected first decision time is~\cite{Weiss83,Yuste96,Schuss19,Holcman19,Lawley20a}:
\begin{equation}
\mathbb{E}[T] \approx \frac{\theta^2}{4\ln{N}}. 
\label{first_decision1}
\end{equation}
 
The mean time decreases logarithmically with $N$, allowing each agent less time to gather private information (Fig.~\ref{Fig:1}e).   When the first decision time, $T$,  is small the remaining beliefs are  distributed almost symmetrically around the origin, and approximately half the population  agrees with the first decision, whether right or wrong. Indeed,  we find that,  
\begin{equation}
\mathbb{E}[a_1|y_1(T)= \pm \theta] \approx \frac{N-1}{2}\Big(1 \pm \frac{\theta}{\sqrt{4\pi \ln{N}}}\Big),
\label{a11}
\end{equation}
where the last term is positive (negative) if the first agent correctly (incorrectly) chooses $H^+$ ($H^-$).
Thus, slightly more than half of a large clique immediately follows a correct first decision, and  
slightly less than half the clique follows a wrong first choice (Fig.~\ref{Fig:1}d shows the mean number and fraction following a correct first decision).
  
The number of agents in excess of half the population following a correct first decision  scales as $N(\ln N)^{-1/2}$. But as the population grows  agents in the first wave accumulate less private information prior to their choice.
We find that  for large $N$, the expected social information communicated by each decision is
\begin{equation} \label{E:update}
\mathbb{E}[R_+(T)] \approx 2 \mathbb{E}[\sqrt{T/\pi}] \approx  \theta /\sqrt{\pi \ln{N}}.
\end{equation}

Thus, as the population increases the size of the first wave, $a_1,$ grows (Fig.~\ref{Fig:1}d), but each first wave decision   
provides less information  (Fig.~\ref{Fig:1}e). However, the logarithmic decrease in 
the revealed information, $R_{\pm}(T)$, is outweighed  by the nearly linear growth in the number of agents, $a_1$: Using 
 Eqs.~(\ref{E:c1}), (\ref{first_decision1}) and \eqref{E:update}, we  find the expected belief update, $ \hat{c}_1^{\pm} \equiv \mathbb{E}[c_1^{\pm}]$,  of undecided agents in the second wave grows nearly linearly in $N$,
\begin{equation}
\hat{c}_1^{\pm} \approx \frac{\theta^2N}{2\pi \ln{N}}.
\label{inc_11}
\end{equation}

The expected update per agent in the second wave following a correct or incorrect decision, $\hat{c}_1^+$  or $\hat{c}_1^-$, is positive: If the first decision 
is correct, more than half the network is in the first wave, and both $(2 \mathbb{E}[a_1] - N  - 2)$  and $R_+(T)$ are positive
in Eq.~\eqref{E:c1}.  Both of these terms are negative when the first decision is wrong.
Thus the second wave is \emph{self-correcting}:   When the network is sufficiently large, $\hat{c}_1^{\pm} > 2\theta$ (see Fig.~\ref{Fig:2}a), and we expect
all undecided agents  to make the correct choice in the second wave, regardless of the choice of the first agent.
As Eq.~\eqref{E:c1} approximates the average belief update, we cannot use it to estimate the probability that the entire clique will decide.  
However, we can use Chebyshev's Inequality to show that when $N \geq 4 \pi (\theta^2(1-x))^{-1}$  the entire network decides by the second wave with probability at least $x$ (See Fig~\ref{Fig:2}b).

{\centering{
\begin{figure}[t!]
\includegraphics[width = \columnwidth]{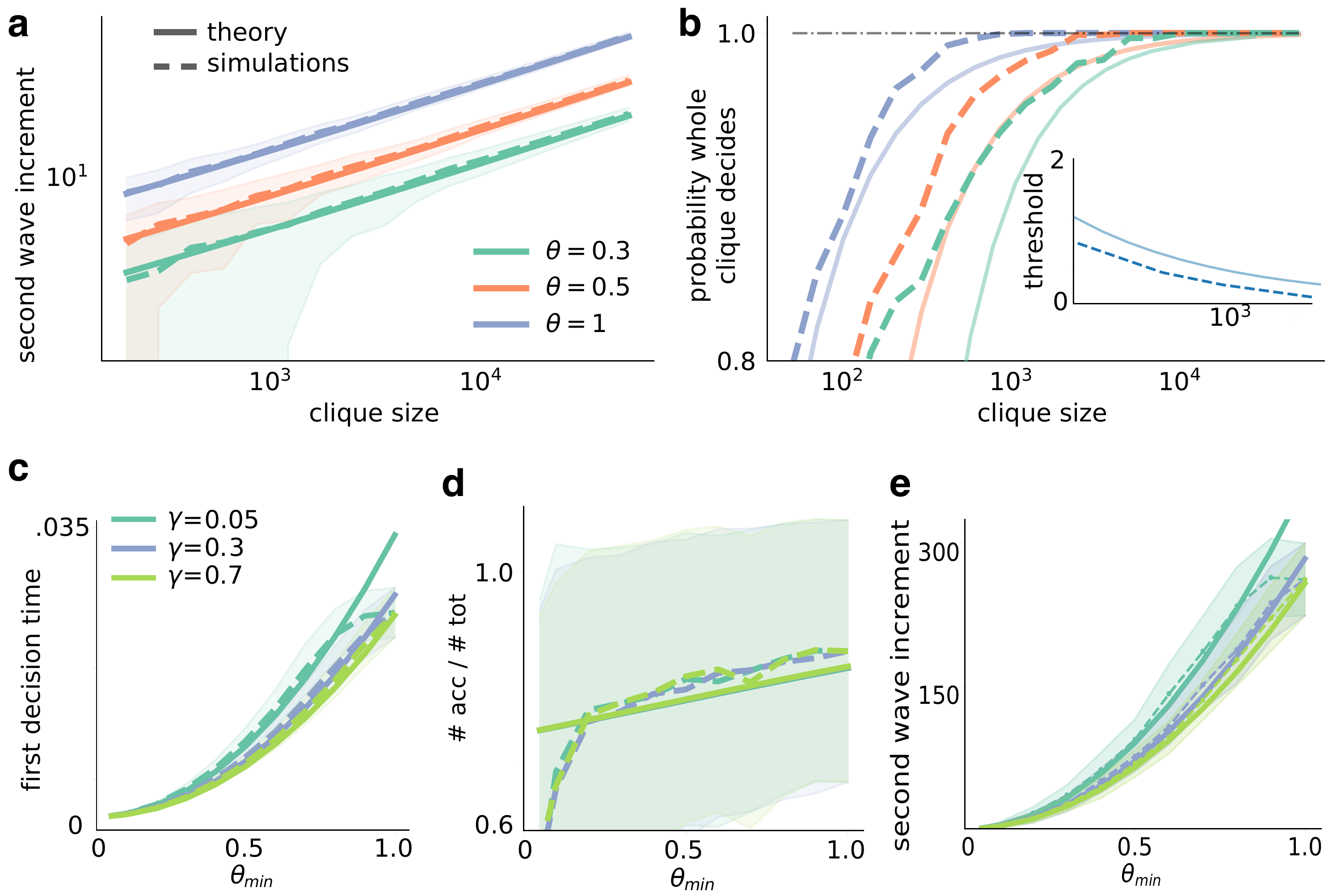}
\vspace{-4mm}
\caption{(Color online) {\bf Decision statistics for large homogeneous  and dichotomous cliques.} (a)~Belief increment $\hat{c}_1^{\pm}$ of agents in the second wave  in homogeneous cliques. (b)~Simulations of the probability the full homogeneous network decides after the second wave as a function of  size, $N$.  Chebyshev's Inequality provides an upper bound on clique size by which the probability is reached.   Inset:  Threshold $\theta$ at which $\hat{c}_1^{\pm} = 2\theta$ as clique size $N$ is varied.  (c) First decision time for dichotomous threshold cliques for various $\gamma$.  (d) Fraction of accurate deciders in dichotomous threshold cliques under consensus bias. (e) Belief increment of agents in the second wave in dichotomous threshold cliques under consensus bias. Clique size $N = 15000$ in panels (c--e).}
\vspace{-4mm}
\label{Fig:2}
\end{figure}
}}

In sum, the first choice triggers a wave of $a_1$ decisions in agreement with the first decider, whether right or wrong.  In large networks, all remaining agents decide correctly in the second wave, regardless of the first agent's decision.  


\begin{figure}
  \includegraphics[width=8cm]{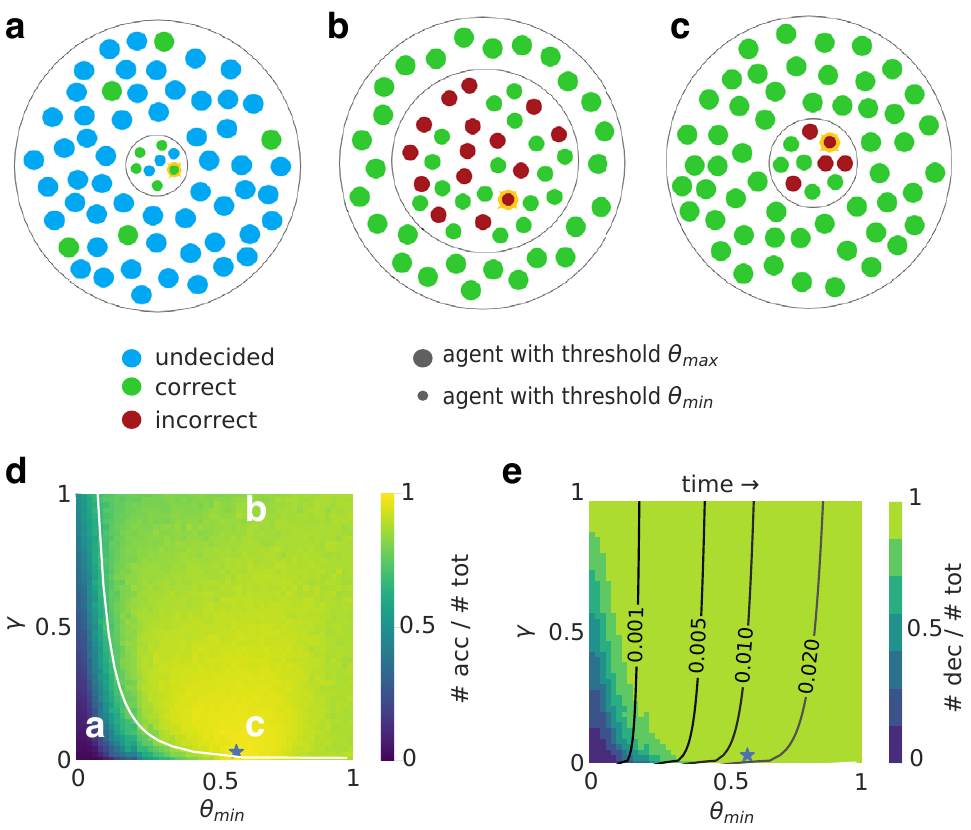}
  \caption{(Color online) {\bf Balancing hasty and deliberate decisions in cliques with dichotomous threshold distributions.} (a)~With too few low threshold agents, the remaining agents do not receive sufficient information to decide after the first wave; (b)~With many low threshold agents, a wrong first decision can sway much of the network; (c)~With the right number of low threshold agents, a small number of hasty agents follows a wrong first decision, but the difference between agreeing and disagreeing low threshold agents is just sufficient for the rest of the clique to choose correctly. (d)~Fraction of the clique choosing accurately for a dichotomous threshold clique. (e)~Fraction of the clique deciding by the end of the second wave.  Isoclines indicate time to first decision. Clique size $N = 15000$ in (b) and (c).
} 

\vspace{-5mm}
  \label{bernoulli}
\end{figure}

\emph{Heterogeneous Populations.}
A population of decision makers is rarely homogeneous.  Some people are quick to make decisions
based on little evidence. Others require substantial information before making a choice \cite{Postmes01,Franks03}.
Some have access to
high quality information, while others rely on poor evidence. How does such diversity impact decisions of the collective?

Here we focus on diversity in the amount of evidence agents require to make a choice by assuming agents'
decision thresholds are distributed on an interval $[\theta_{\rm min},\theta_{\rm max}]$.  Agents with a  low threshold are more likely to decide first, but also to make a wrong choice \cite{Srivastava14}.  The ensuing exchange of social information depends on assumptions agents make about each other. 
While collective decisions in heterogeneous networks under consensus bias are similar to those in homogeneous populations, omniscient agents can leverage quick, unreliable decisions to improve the response of the population.


\emph{a. Dichotomous threshold distribution.}  The case when all agents have either a high or a low threshold is tractable and sheds light on more general examples. Before a decision the belief of each agent  evolves according to Eq.~(\ref{langevin}) on a symmetric interval with absorbing boundaries at $-\theta_i < 0 < \theta_i$. We assume that  $\gamma N$ agents share threshold $\theta_{\rm min}$ and $(1-\gamma)N$ share threshold $\theta_{\rm max}$ for some $0< \theta_{\rm min} < \theta_{\rm max}$ and $\gamma \in (0,1)$. The first decision is then most likely made by an agent with a low threshold, and is thus fast but unreliable.  We therefore use the approximation  $\mathbb{E}[T] \approx \frac{\theta_{\rm min}^2}{4\ln{(\gamma N)}}$ which breaks down when $0<\gamma \ll 1$ but works well otherwise (see Fig.~\ref{Fig:2}c).  The probability that this decision is correct is $(1 + \exp(-\theta_{\rm min}))^{-1}$~\cite{Bogacz2006}.  
The social network is homogeneous from the perspective of an observer under consensus bias. We thus again expect about half of the  network to 
follow the first choice, whether right or wrong. Indeed, the expected size of the first wave is given by an expression similar to Eq.~\eqref{a11},  $\mathbb{E}[a_1] \approx \frac{N-1}{2}\Big(1 \pm \frac{\theta_{\rm min}}{\sqrt{4\pi \ln{\gamma N}}}\Big)$.  
The expected belief increment in the second wave is $\hat{c}_1^{\pm} \approx \frac{\theta_{\rm min}^2 \gamma N}{2\pi \ln{\gamma N}}$ which is analogous to Eq.~\eqref{inc_11}, and is governed primarily by the timing of the first choice  (See Fig.~\ref{Fig:2}e).
 In large populations decisions happen quickly, before the belief distributions can interact with the boundaries. Therefore the expected belief increment, $\hat{c}_1^{\pm},$  is~\emph{approximately independent of 
the observer's threshold}:   Following the first wave low and high threshold agents make the same update.

As in homogeneous networks, the size of the expected update, $\hat{c}_1^{\pm}$, grows with population size. When the update exceeds $2\theta_{\rm max}$, we expect all agents to decide by the second wave.  If the first decision is correct, the entire clique follows. A wrong first choice is followed by the first wave constituting about half the network (see Fig.~\ref{Fig:2}d), while the second wave decides correctly. Under consensus bias, dichotomous cliques behave as if they were homogenous with threshold $\theta_{\rm min}$: Uninformed agents govern decisions, leading to fast, inaccurate choices.


In contrast, \emph{omniscient} agents correctly weigh evidence revealed by a hasty first decider. We expect about half of the low threshold agents, $\gamma N/2$, to decide in the first wave. Indeed, we find $\mathbb{E}[a_1] \approx \frac{\gamma N-1}{2}\Big(1 \pm \frac{\theta_{\rm min}}{\sqrt{4\pi \ln{\gamma N}}}\Big)$.
The evidence revealed by a single low threshold agent is unlikely to sway high threshold agents. However, if the subpopulation of low threshold agents is sufficiently large, the difference between
those convinced and unconvinced by the first choice can trigger a correct decision in the rest of the population (See Fig.~\ref{bernoulli}b,c).

Thus, in a network of omniscient agents, hasty observers again govern the speed of the first decision and mostly comprise the first wave.  The remaining agents can then observe the choices of the early adopters to make the right decision.  The fraction of wrong decisions can thus be smaller than in homogeneous networks.  

In finite populations this argument requires $\gamma$ and $\theta_{\rm min}$ be large enough for the first wave to convince the remainder of the population (Fig.~\ref{bernoulli}a), but small enough to buffer the majority of the population from following an incorrect first choice (Fig.~\ref{bernoulli}b). We thus expect that the population makes the best decisions at intermediate values of $\gamma$ and $\theta_{\rm min}$ (star in Fig.~\ref{bernoulli}d). A balance between these cases is reached when $\hat{c}_1^{-} = 2 \theta_{\rm max}$, which corresponds to a fraction of low threshold agents given by 
$$
\gamma \approx \frac{4\pi \theta_{\rm max}}{N} \frac{\ln{N}}{\theta_{\rm min}^2}.
$$
Almost all agents decide by the second wave (Fig.~\ref{bernoulli}e).

\begin{figure}[t!]
  \includegraphics[width=\columnwidth]{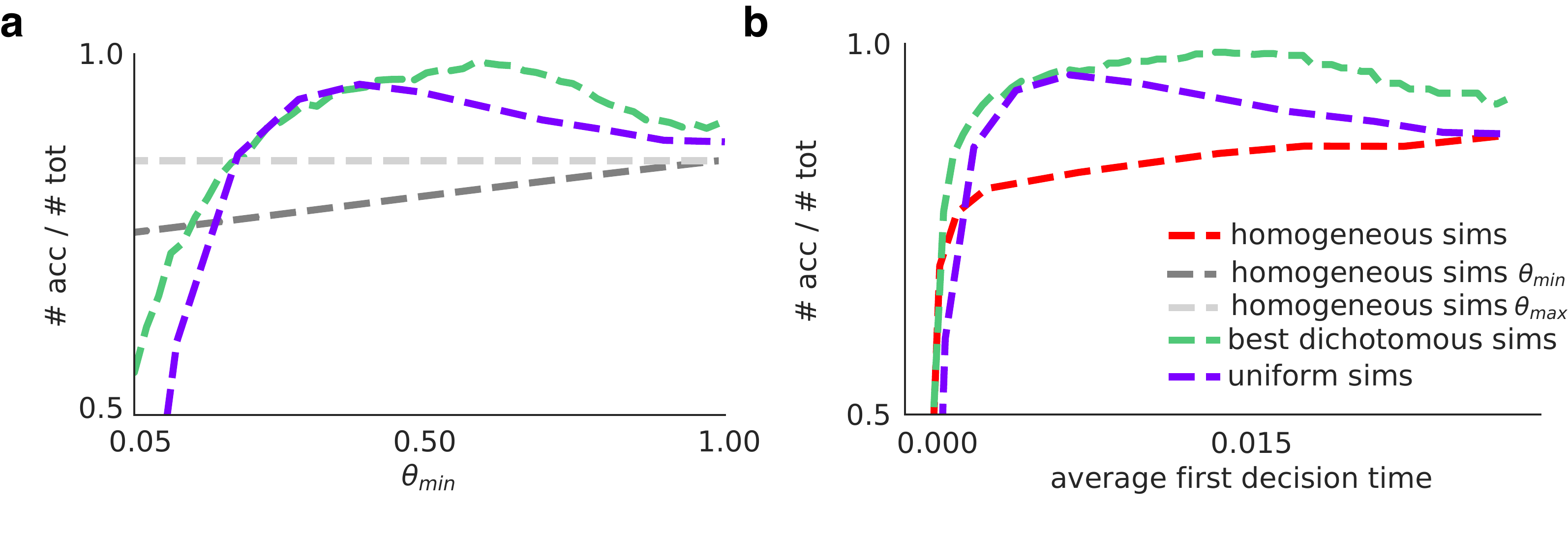}
  \vspace{-7mm}
  \caption{(Color online) {\bf Improved accuracy for fixed decision speed.} (a) Mean fraction of the entire clique choosing accurately after two waves for different threshold distributions in an omniscient population.  For the dichotomous case, $\gamma$ is chosen to maximize accuracy for each $\theta_{\rm min}$ value. (b) Over a  range of possible first decision times, heterogeneous cliques give better accuracy than homogeneous ones with omniscient social updating. Here, $N = 15000$ and $\theta_{\rm max} = 1$. } 
  \vspace{-4mm}
  \label{socialExplanation}
\end{figure}

Thus a finite population with dichotomous thresholds can sacrifice a small fraction of early adopters so most of the population makes a fast, correct choice.  Agents in heterogenous networks can thus decide more quickly, and outperform agents in homogeneous networks in recovering from a wrong first choice (Figs.~\ref{bernoulli}c and \ref{socialExplanation}).  

\emph{b. Different threshold distributions.}  We next consider $N$ agents with thresholds, $\theta_i,$ following different distributions supported on the interval $[\theta_{\rm min},\theta_{\rm max}]$. The expected time to the first decision is again governed by the smallest threshold, $\theta_{\rm min}$, and  under consensus biased the size of the first wave is 
$
\mathbb{E}[a_1] \approx \frac{N-1}{2}\Big[1 \pm \frac{\theta_{\rm min}}{\sqrt{4\pi \ln{N}}}\Big],
$
with the sign determined by the first decision.  In either case, the increment to the undecided agents after the first wave is given by Eq.~(\ref{inc_11}) with $\theta_{\rm min}$ replacing $\theta$.
For sufficiently large $N$, $\hat{c}_1 > 2\theta_{\rm max}$.  Therefore, under consensus bias the clique again behaves as a homogeneous clique with 
threshold $\theta_{\rm min}$.

The omniscient case is more complicated.  Numerical simulations show the trends observed in the dichotomous case persist for a large class of threshold distributions: Hasty agents decide first, and deliberate agents decide based on which early adopters followed the first choice (See Figs.~\ref{socialExplanation}), leading to faster and more accurate choices than in homogeneous networks.

\emph{Conclusion.} Our model of collective decision making is analytically tractable and shows how diverse populations can make quicker, more accurate decisions than homogeneous ones.  Previous models of collective decision-making have either ignored temporal aspects of evidence accumulation~\cite{Goyal2012,mueller2013,Mann18} or did not describe rational agents~\cite{Watts2002,Caginalp2017}.  Our model incorporates both aspects and can serve as a baseline to understand when and how decisions depart from rationality~\cite{geisler03}.

There are a number of ways to include more realistic features in the model:  Observations can be correlated, rather than \textit{conditionally independent},  resulting in correlated noise in Eqs.~(\ref{langevin}) and (\ref{smoluchowski})~\cite{moreno10}.  Agents could accumulate evidence at different rates, resulting in each having different drift and diffusion coefficients~\cite{Bogacz2006,veliz16}.  The framework we provide can be extended to these cases to probe how different conditions influence decisions.\\
\begin{acknowledgements}
\vspace{-3mm}
We would like to thank Thibaud Taillefumier for helpful comments. This work was supported by NSF/NIH CRCNS grant R01MH115557. BK and KJ were supported by NSF grant DMS-1662305. KJ was also supported by NSF NeuroNex grant DBI-1707400. ZPK was also supported by NSF grant DMS-1853630.
WO was supported by NSF grant DMS-1816315 and NIH grant R01GM117138.
\end{acknowledgements}

\bibliography{references}

\begin{thebibliography}{36}%
\makeatletter
\providecommand \@ifxundefined [1]{%
 \@ifx{#1\undefined}
}%
\providecommand \@ifnum [1]{%
 \ifnum #1\expandafter \@firstoftwo
 \else \expandafter \@secondoftwo
 \fi
}%
\providecommand \@ifx [1]{%
 \ifx #1\expandafter \@firstoftwo
 \else \expandafter \@secondoftwo
 \fi
}%
\providecommand \natexlab [1]{#1}%
\providecommand \enquote  [1]{``#1''}%
\providecommand \bibnamefont  [1]{#1}%
\providecommand \bibfnamefont [1]{#1}%
\providecommand \citenamefont [1]{#1}%
\providecommand \href@noop [0]{\@secondoftwo}%
\providecommand \href [0]{\begingroup \@sanitize@url \@href}%
\providecommand \@href[1]{\@@startlink{#1}\@@href}%
\providecommand \@@href[1]{\endgroup#1\@@endlink}%
\providecommand \@sanitize@url [0]{\catcode `\\12\catcode `\$12\catcode
  `\&12\catcode `\#12\catcode `\^12\catcode `\_12\catcode `\%12\relax}%
\providecommand \@@startlink[1]{}%
\providecommand \@@endlink[0]{}%
\providecommand \url  [0]{\begingroup\@sanitize@url \@url }%
\providecommand \@url [1]{\endgroup\@href {#1}{\urlprefix }}%
\providecommand \urlprefix  [0]{URL }%
\providecommand \Eprint [0]{\href }%
\providecommand \doibase [0]{http://dx.doi.org/}%
\providecommand \selectlanguage [0]{\@gobble}%
\providecommand \bibinfo  [0]{\@secondoftwo}%
\providecommand \bibfield  [0]{\@secondoftwo}%
\providecommand \translation [1]{[#1]}%
\providecommand \BibitemOpen [0]{}%
\providecommand \bibitemStop [0]{}%
\providecommand \bibitemNoStop [0]{.\EOS\space}%
\providecommand \EOS [0]{\spacefactor3000\relax}%
\providecommand \BibitemShut  [1]{\csname bibitem#1\endcsname}%
\let\auto@bib@innerbib\@empty
\bibitem [{\citenamefont {Couzin}(2009)}]{Couzin2009}%
  \BibitemOpen
  \bibfield  {author} {\bibinfo {author} {\bibfnamefont {I.~D.}\ \bibnamefont
  {Couzin}},\ }\href@noop {} {\bibfield  {journal} {\bibinfo  {journal} {Trends
  in cognitive sciences}\ }\textbf {\bibinfo {volume} {13}},\ \bibinfo {pages}
  {36} (\bibinfo {year} {2009})}\BibitemShut {NoStop}%
\bibitem [{\citenamefont {Edwards}(1954)}]{Edwards54}%
  \BibitemOpen
  \bibfield  {author} {\bibinfo {author} {\bibfnamefont {W.}~\bibnamefont
  {Edwards}},\ }\href@noop {} {\bibfield  {journal} {\bibinfo  {journal}
  {Psychological Bulletin}\ }\textbf {\bibinfo {volume} {51}},\ \bibinfo
  {pages} {380} (\bibinfo {year} {1954})}\BibitemShut {NoStop}%
\bibitem [{\citenamefont {Frith}\ and\ \citenamefont
  {Frith}(2012)}]{Frith2012}%
  \BibitemOpen
  \bibfield  {author} {\bibinfo {author} {\bibfnamefont {C.~D.}\ \bibnamefont
  {Frith}}\ and\ \bibinfo {author} {\bibfnamefont {U.}~\bibnamefont {Frith}},\
  }\href@noop {} {\bibfield  {journal} {\bibinfo  {journal} {Annual review of
  psychology}\ }\textbf {\bibinfo {volume} {63}},\ \bibinfo {pages} {287}
  (\bibinfo {year} {2012})}\BibitemShut {NoStop}%
\bibitem [{\citenamefont {P\'{e}rez-Escudero}\ and\ \citenamefont
  {Polavieja}(2011)}]{PE2011}%
  \BibitemOpen
  \bibfield  {author} {\bibinfo {author} {\bibfnamefont {A.}~\bibnamefont
  {P\'{e}rez-Escudero}}\ and\ \bibinfo {author} {\bibfnamefont {G.~G.~D.}\
  \bibnamefont {Polavieja}},\ }\href@noop {} {\bibfield  {journal} {\bibinfo
  {journal} {PLoS Computational Biology}\ } (\bibinfo {year}
  {2011})}\BibitemShut {NoStop}%
\bibitem [{\citenamefont {Arganda}\ \emph {et~al.}(2012)\citenamefont
  {Arganda}, \citenamefont {P\'{e}rez-Escudero},\ and\ \citenamefont
  {Polavieja}}]{Arganda12}%
  \BibitemOpen
  \bibfield  {author} {\bibinfo {author} {\bibfnamefont {S.}~\bibnamefont
  {Arganda}}, \bibinfo {author} {\bibfnamefont {A.}~\bibnamefont
  {P\'{e}rez-Escudero}}, \ and\ \bibinfo {author} {\bibfnamefont {G.~G.~D.}\
  \bibnamefont {Polavieja}},\ }\href@noop {} {\bibfield  {journal} {\bibinfo
  {journal} {Proceedings of the National Academy of Sciences}\ }\textbf
  {\bibinfo {volume} {109}},\ \bibinfo {pages} {20508} (\bibinfo {year}
  {2012})}\BibitemShut {NoStop}%
\bibitem [{\citenamefont {Mann}(2018)}]{Mann18}%
  \BibitemOpen
  \bibfield  {author} {\bibinfo {author} {\bibfnamefont {R.~P.}\ \bibnamefont
  {Mann}},\ }\href@noop {} {\bibfield  {journal} {\bibinfo  {journal}
  {Proceedings of the National Academy of Sciences}\ }\textbf {\bibinfo
  {volume} {115}},\ \bibinfo {pages} {E10387} (\bibinfo {year}
  {2018})}\BibitemShut {NoStop}%
\bibitem [{\citenamefont {Mann}(2020)}]{Mann2020}%
  \BibitemOpen
  \bibfield  {author} {\bibinfo {author} {\bibfnamefont {R.~P.}\ \bibnamefont
  {Mann}},\ }\href@noop {} {\bibfield  {journal} {\bibinfo  {journal}
  {Proceedings of the National Academy of Sciences}\ } (\bibinfo {year}
  {2020})}\BibitemShut {NoStop}%
\bibitem [{\citenamefont {Ward}\ \emph {et~al.}(2008)\citenamefont {Ward},
  \citenamefont {Sumpter}, \citenamefont {Couzin}, \citenamefont {Hart},\ and\
  \citenamefont {Krause}}]{Ward08}%
  \BibitemOpen
  \bibfield  {author} {\bibinfo {author} {\bibfnamefont {A.~J.~W.}\
  \bibnamefont {Ward}}, \bibinfo {author} {\bibfnamefont {D.~J.~T.}\
  \bibnamefont {Sumpter}}, \bibinfo {author} {\bibfnamefont {I.~D.}\
  \bibnamefont {Couzin}}, \bibinfo {author} {\bibfnamefont {P.~J.~B.}\
  \bibnamefont {Hart}}, \ and\ \bibinfo {author} {\bibfnamefont
  {J.}~\bibnamefont {Krause}},\ }\href@noop {} {\bibfield  {journal} {\bibinfo
  {journal} {Proceedings of the National Academy of Sciences}\ }\textbf
  {\bibinfo {volume} {105}},\ \bibinfo {pages} {6948} (\bibinfo {year}
  {2008})}\BibitemShut {NoStop}%
\bibitem [{\citenamefont {Ward}\ \emph {et~al.}(2012)\citenamefont {Ward},
  \citenamefont {Krause},\ and\ \citenamefont {Sumpter}}]{Ward12}%
  \BibitemOpen
  \bibfield  {author} {\bibinfo {author} {\bibfnamefont {A.}~\bibnamefont
  {Ward}}, \bibinfo {author} {\bibfnamefont {J.}~\bibnamefont {Krause}}, \ and\
  \bibinfo {author} {\bibfnamefont {D.}~\bibnamefont {Sumpter}},\ }\href@noop
  {} {\bibfield  {journal} {\bibinfo  {journal} {PloS one}\ }\textbf {\bibinfo
  {volume} {7}} (\bibinfo {year} {2012})}\BibitemShut {NoStop}%
\bibitem [{\citenamefont {Gall}\ \emph {et~al.}(2017)\citenamefont {Gall},
  \citenamefont {Strandburg-Peshkin}, \citenamefont {Clutton-Brock},\ and\
  \citenamefont {Manser}}]{Gall17}%
  \BibitemOpen
  \bibfield  {author} {\bibinfo {author} {\bibfnamefont {G.~E.}\ \bibnamefont
  {Gall}}, \bibinfo {author} {\bibfnamefont {A.}~\bibnamefont
  {Strandburg-Peshkin}}, \bibinfo {author} {\bibfnamefont {T.}~\bibnamefont
  {Clutton-Brock}}, \ and\ \bibinfo {author} {\bibfnamefont {M.~B.}\
  \bibnamefont {Manser}},\ }\href@noop {} {\bibfield  {journal} {\bibinfo
  {journal} {Animal Behavior}\ }\textbf {\bibinfo {volume} {132}},\ \bibinfo
  {pages} {91} (\bibinfo {year} {2017})}\BibitemShut {NoStop}%
\bibitem [{\citenamefont {Strandburg-Peshkin}\ \emph
  {et~al.}(2015)\citenamefont {Strandburg-Peshkin}, \citenamefont {Farine},
  \citenamefont {Couzin},\ and\ \citenamefont {Crofoot}}]{SP15}%
  \BibitemOpen
  \bibfield  {author} {\bibinfo {author} {\bibfnamefont {A.}~\bibnamefont
  {Strandburg-Peshkin}}, \bibinfo {author} {\bibfnamefont {D.~R.}\ \bibnamefont
  {Farine}}, \bibinfo {author} {\bibfnamefont {I.~D.}\ \bibnamefont {Couzin}},
  \ and\ \bibinfo {author} {\bibfnamefont {M.~C.}\ \bibnamefont {Crofoot}},\
  }\href@noop {} {\bibfield  {journal} {\bibinfo  {journal} {Science}\ }\textbf
  {\bibinfo {volume} {348}},\ \bibinfo {pages} {1358} (\bibinfo {year}
  {2015})}\BibitemShut {NoStop}%
\bibitem [{\citenamefont {Perna}\ \emph {et~al.}(2012)\citenamefont {Perna},
  \citenamefont {Granovskiy}, \citenamefont {Garnier}, \citenamefont {Nicolis},
  \citenamefont {Lab\'{e}dan}, \citenamefont {Theraulaz}, \citenamefont
  {Fourcassi\'{e}},\ and\ \citenamefont {Sumpter}}]{Perna12}%
  \BibitemOpen
  \bibfield  {author} {\bibinfo {author} {\bibfnamefont {A.}~\bibnamefont
  {Perna}}, \bibinfo {author} {\bibfnamefont {B.}~\bibnamefont {Granovskiy}},
  \bibinfo {author} {\bibfnamefont {S.}~\bibnamefont {Garnier}}, \bibinfo
  {author} {\bibfnamefont {S.~C.}\ \bibnamefont {Nicolis}}, \bibinfo {author}
  {\bibfnamefont {M.}~\bibnamefont {Lab\'{e}dan}}, \bibinfo {author}
  {\bibfnamefont {G.}~\bibnamefont {Theraulaz}}, \bibinfo {author}
  {\bibfnamefont {V.}~\bibnamefont {Fourcassi\'{e}}}, \ and\ \bibinfo {author}
  {\bibfnamefont {D.~J.~T.}\ \bibnamefont {Sumpter}},\ }\href@noop {}
  {\bibfield  {journal} {\bibinfo  {journal} {PLoS Computational Biology}\
  }\textbf {\bibinfo {volume} {8}},\ \bibinfo {pages} {e1002592} (\bibinfo
  {year} {2012})}\BibitemShut {NoStop}%
\bibitem [{\citenamefont {Walker}\ \emph {et~al.}(2017)\citenamefont {Walker},
  \citenamefont {King}, \citenamefont {McNutt},\ and\ \citenamefont
  {Jordan}}]{Walker17}%
  \BibitemOpen
  \bibfield  {author} {\bibinfo {author} {\bibfnamefont {R.~H.}\ \bibnamefont
  {Walker}}, \bibinfo {author} {\bibfnamefont {A.~J.}\ \bibnamefont {King}},
  \bibinfo {author} {\bibfnamefont {J.~W.}\ \bibnamefont {McNutt}}, \ and\
  \bibinfo {author} {\bibfnamefont {N.~R.}\ \bibnamefont {Jordan}},\
  }\href@noop {} {\bibfield  {journal} {\bibinfo  {journal} {Prceedings of the
  Royal Society B}\ }\textbf {\bibinfo {volume} {284}} (\bibinfo {year}
  {2017})}\BibitemShut {NoStop}%
\bibitem [{\citenamefont {Faria}\ \emph {et~al.}(2010)\citenamefont {Faria},
  \citenamefont {Krause},\ and\ \citenamefont {Krause}}]{Faria10}%
  \BibitemOpen
  \bibfield  {author} {\bibinfo {author} {\bibfnamefont {J.}~\bibnamefont
  {Faria}}, \bibinfo {author} {\bibfnamefont {S.}~\bibnamefont {Krause}}, \
  and\ \bibinfo {author} {\bibfnamefont {J.}~\bibnamefont {Krause}},\
  }\href@noop {} {\bibfield  {journal} {\bibinfo  {journal} {Behavioral
  Ecology}\ }\textbf {\bibinfo {volume} {21}},\ \bibinfo {pages} {1236}
  (\bibinfo {year} {2010})}\BibitemShut {NoStop}%
\bibitem [{\citenamefont {Ratcliff}(1978)}]{Ratcliff1978theory}%
  \BibitemOpen
  \bibfield  {author} {\bibinfo {author} {\bibfnamefont {R.}~\bibnamefont
  {Ratcliff}},\ }\href@noop {} {\bibfield  {journal} {\bibinfo  {journal}
  {Psychological review}\ }\textbf {\bibinfo {volume} {85}},\ \bibinfo {pages}
  {59} (\bibinfo {year} {1978})}\BibitemShut {NoStop}%
\bibitem [{\citenamefont {Gold}\ and\ \citenamefont {Shadlen}(2002)}]{Gold02}%
  \BibitemOpen
  \bibfield  {author} {\bibinfo {author} {\bibfnamefont {J.~I.}\ \bibnamefont
  {Gold}}\ and\ \bibinfo {author} {\bibfnamefont {M.~N.}\ \bibnamefont
  {Shadlen}},\ }\href@noop {} {\bibfield  {journal} {\bibinfo  {journal}
  {Neuron}\ }\textbf {\bibinfo {volume} {36}},\ \bibinfo {pages} {299}
  (\bibinfo {year} {2002})}\BibitemShut {NoStop}%
\bibitem [{\citenamefont {Bogacz}\ \emph {et~al.}(2006)\citenamefont {Bogacz},
  \citenamefont {Brown}, \citenamefont {Moehlis}, \citenamefont {Holmes},\ and\
  \citenamefont {Cohen}}]{Bogacz2006}%
  \BibitemOpen
  \bibfield  {author} {\bibinfo {author} {\bibfnamefont {R.}~\bibnamefont
  {Bogacz}}, \bibinfo {author} {\bibfnamefont {E.}~\bibnamefont {Brown}},
  \bibinfo {author} {\bibfnamefont {J.}~\bibnamefont {Moehlis}}, \bibinfo
  {author} {\bibfnamefont {P.}~\bibnamefont {Holmes}}, \ and\ \bibinfo {author}
  {\bibfnamefont {J.~D.}\ \bibnamefont {Cohen}},\ }\href@noop {} {\bibfield
  {journal} {\bibinfo  {journal} {Psychological Review}\ }\textbf {\bibinfo
  {volume} {113}},\ \bibinfo {pages} {700} (\bibinfo {year}
  {2006})}\BibitemShut {NoStop}%
\bibitem [{\citenamefont {Karamched}\ \emph {et~al.}(2020)\citenamefont
  {Karamched}, \citenamefont {Stolarczyk}, \citenamefont {Kilpatrick},\ and\
  \citenamefont {Josi\'{c}}}]{Karamched20}%
  \BibitemOpen
  \bibfield  {author} {\bibinfo {author} {\bibfnamefont {B.~R.}\ \bibnamefont
  {Karamched}}, \bibinfo {author} {\bibfnamefont {S.}~\bibnamefont
  {Stolarczyk}}, \bibinfo {author} {\bibfnamefont {Z.~P.}\ \bibnamefont
  {Kilpatrick}}, \ and\ \bibinfo {author} {\bibfnamefont {K.}~\bibnamefont
  {Josi\'{c}}},\ }\href@noop {} {\bibfield  {journal} {\bibinfo  {journal}
  {SIAM Journal on Applied Dynamical Systems}\ }\textbf {\bibinfo {volume} {in
  press}} (\bibinfo {year} {2020})}\BibitemShut {NoStop}%
\bibitem [{\citenamefont {Goyal}(2012)}]{Goyal2012}%
  \BibitemOpen
  \bibfield  {author} {\bibinfo {author} {\bibfnamefont {S.~b.}\ \bibnamefont
  {Goyal}},\ }\href {\doibase 10.1007/s00712-008-0036-9} {\emph {\bibinfo
  {title} {Connections: An Introduction to the Economics of Networks}}}\
  (\bibinfo  {publisher} {Princeton University Press},\ \bibinfo {year}
  {2012})\ pp.\ \bibinfo {pages} {1--289},\ \Eprint
  {http://arxiv.org/abs/9809069v1} {arXiv:9809069v1 [arXiv:gr-qc]} \BibitemShut
  {NoStop}%
\bibitem [{\citenamefont {Veliz-Cuba}\ \emph {et~al.}(2016)\citenamefont
  {Veliz-Cuba}, \citenamefont {Kilpatrick},\ and\ \citenamefont
  {Josi{\'c}}}]{veliz16}%
  \BibitemOpen
  \bibfield  {author} {\bibinfo {author} {\bibfnamefont {A.}~\bibnamefont
  {Veliz-Cuba}}, \bibinfo {author} {\bibfnamefont {Z.~P.}\ \bibnamefont
  {Kilpatrick}}, \ and\ \bibinfo {author} {\bibfnamefont {K.}~\bibnamefont
  {Josi{\'c}}},\ }\href@noop {} {\bibfield  {journal} {\bibinfo  {journal}
  {SIAM Review}\ }\textbf {\bibinfo {volume} {58}},\ \bibinfo {pages} {264}
  (\bibinfo {year} {2016})}\BibitemShut {NoStop}%
\bibitem [{Note1()}]{Note1}%
  \BibitemOpen
  \bibinfo {note} {Supplementary Material contains details about the
  simulations and calculations.}\BibitemShut {Stop}%
\bibitem [{\citenamefont {Cox}\ and\ \citenamefont {Miller}(1965)}]{cox65}%
  \BibitemOpen
  \bibfield  {author} {\bibinfo {author} {\bibfnamefont {D.~R.}\ \bibnamefont
  {Cox}}\ and\ \bibinfo {author} {\bibfnamefont {H.~D.}\ \bibnamefont
  {Miller}},\ }\href@noop {} {\emph {\bibinfo {title} {The theory of stochastic
  processes}}}\ (\bibinfo  {publisher} {Chapman and Hall},\ \bibinfo {year}
  {1965})\BibitemShut {NoStop}%
\bibitem [{\citenamefont {Drugowitsch}(2016)}]{Drugowitsch16}%
  \BibitemOpen
  \bibfield  {author} {\bibinfo {author} {\bibfnamefont {J.}~\bibnamefont
  {Drugowitsch}},\ }\href@noop {} {\bibfield  {journal} {\bibinfo  {journal}
  {Scientific Reports}\ }\textbf {\bibinfo {volume} {6}},\ \bibinfo {pages}
  {20490} (\bibinfo {year} {2016})}\BibitemShut {NoStop}%
\bibitem [{\citenamefont {Weiss}\ \emph {et~al.}(1983)\citenamefont {Weiss},
  \citenamefont {Shuler},\ and\ \citenamefont {Lindenberg}}]{Weiss83}%
  \BibitemOpen
  \bibfield  {author} {\bibinfo {author} {\bibfnamefont {G.~H.}\ \bibnamefont
  {Weiss}}, \bibinfo {author} {\bibfnamefont {K.~E.}\ \bibnamefont {Shuler}}, \
  and\ \bibinfo {author} {\bibfnamefont {K.}~\bibnamefont {Lindenberg}},\
  }\href@noop {} {\bibfield  {journal} {\bibinfo  {journal} {J. Stat. Phys.}\
  }\textbf {\bibinfo {volume} {31}},\ \bibinfo {pages} {255} (\bibinfo {year}
  {1983})}\BibitemShut {NoStop}%
\bibitem [{\citenamefont {Yuste}\ and\ \citenamefont
  {Lindenberg}(1996)}]{Yuste96}%
  \BibitemOpen
  \bibfield  {author} {\bibinfo {author} {\bibfnamefont {S.~B.}\ \bibnamefont
  {Yuste}}\ and\ \bibinfo {author} {\bibfnamefont {K.}~\bibnamefont
  {Lindenberg}},\ }\href@noop {} {\bibfield  {journal} {\bibinfo  {journal} {J.
  Stat. Phys.}\ }\textbf {\bibinfo {volume} {85}},\ \bibinfo {pages} {501}
  (\bibinfo {year} {1996})}\BibitemShut {NoStop}%
\bibitem [{\citenamefont {Schuss}\ \emph {et~al.}(2019)\citenamefont {Schuss},
  \citenamefont {Basnayake},\ and\ \citenamefont {Holcman}}]{Schuss19}%
  \BibitemOpen
  \bibfield  {author} {\bibinfo {author} {\bibfnamefont {Z.}~\bibnamefont
  {Schuss}}, \bibinfo {author} {\bibfnamefont {K.}~\bibnamefont {Basnayake}}, \
  and\ \bibinfo {author} {\bibfnamefont {D.}~\bibnamefont {Holcman}},\
  }\href@noop {} {\bibfield  {journal} {\bibinfo  {journal} {Physics of Life
  Reviews}\ } (\bibinfo {year} {2019})}\BibitemShut {NoStop}%
\bibitem [{\citenamefont {Basnayake}\ \emph {et~al.}(2019)\citenamefont
  {Basnayake}, \citenamefont {Schuss},\ and\ \citenamefont
  {Holcman}}]{Holcman19}%
  \BibitemOpen
  \bibfield  {author} {\bibinfo {author} {\bibfnamefont {K.}~\bibnamefont
  {Basnayake}}, \bibinfo {author} {\bibfnamefont {Z.}~\bibnamefont {Schuss}}, \
  and\ \bibinfo {author} {\bibfnamefont {D.}~\bibnamefont {Holcman}},\
  }\href@noop {} {\bibfield  {journal} {\bibinfo  {journal} {Journal of
  Nonlinear Science}\ }\textbf {\bibinfo {volume} {29}},\ \bibinfo {pages}
  {461} (\bibinfo {year} {2019})}\BibitemShut {NoStop}%
\bibitem [{\citenamefont {Lawley}\ and\ \citenamefont
  {Madrid}(2020)}]{Lawley20a}%
  \BibitemOpen
  \bibfield  {author} {\bibinfo {author} {\bibfnamefont {S.~D.}\ \bibnamefont
  {Lawley}}\ and\ \bibinfo {author} {\bibfnamefont {J.~B.}\ \bibnamefont
  {Madrid}},\ }\href@noop {} {\bibfield  {journal} {\bibinfo  {journal}
  {Journal of Nonlinear Science}\ ,\ \bibinfo {pages} {1}} (\bibinfo {year}
  {2020})}\BibitemShut {NoStop}%
\bibitem [{\citenamefont {Postmes}\ \emph {et~al.}(2001)\citenamefont
  {Postmes}, \citenamefont {Spears},\ and\ \citenamefont
  {Cihangir}}]{Postmes01}%
  \BibitemOpen
  \bibfield  {author} {\bibinfo {author} {\bibfnamefont {T.}~\bibnamefont
  {Postmes}}, \bibinfo {author} {\bibfnamefont {R.}~\bibnamefont {Spears}}, \
  and\ \bibinfo {author} {\bibfnamefont {S.}~\bibnamefont {Cihangir}},\
  }\href@noop {} {\bibfield  {journal} {\bibinfo  {journal} {Journal of
  Personality and Social Psychology}\ }\textbf {\bibinfo {volume} {80}},\
  \bibinfo {pages} {918} (\bibinfo {year} {2001})}\BibitemShut {NoStop}%
\bibitem [{\citenamefont {Franks}\ \emph {et~al.}(2003)\citenamefont {Franks},
  \citenamefont {Dornhaus}, \citenamefont {Fitzsimmons},\ and\ \citenamefont
  {Stevens}}]{Franks03}%
  \BibitemOpen
  \bibfield  {author} {\bibinfo {author} {\bibfnamefont {N.~R.}\ \bibnamefont
  {Franks}}, \bibinfo {author} {\bibfnamefont {A.}~\bibnamefont {Dornhaus}},
  \bibinfo {author} {\bibfnamefont {J.~P.}\ \bibnamefont {Fitzsimmons}}, \ and\
  \bibinfo {author} {\bibfnamefont {M.}~\bibnamefont {Stevens}},\ }\href@noop
  {} {\bibfield  {journal} {\bibinfo  {journal} {Prceedings of the Royal
  Society B}\ }\textbf {\bibinfo {volume} {270}},\ \bibinfo {pages} {2457}
  (\bibinfo {year} {2003})}\BibitemShut {NoStop}%
\bibitem [{\citenamefont {Srivastava}\ and\ \citenamefont
  {Leonard}(2014)}]{Srivastava14}%
  \BibitemOpen
  \bibfield  {author} {\bibinfo {author} {\bibfnamefont {V.}~\bibnamefont
  {Srivastava}}\ and\ \bibinfo {author} {\bibfnamefont {N.~E.}\ \bibnamefont
  {Leonard}},\ }\href@noop {} {\bibfield  {journal} {\bibinfo  {journal} {IEEE
  Transactions on Control of Network Systems}\ }\textbf {\bibinfo {volume}
  {1}},\ \bibinfo {pages} {121} (\bibinfo {year} {2014})}\BibitemShut {NoStop}%
\bibitem [{\citenamefont {Mueller-Frank}(2013)}]{mueller2013}%
  \BibitemOpen
  \bibfield  {author} {\bibinfo {author} {\bibfnamefont {M.}~\bibnamefont
  {Mueller-Frank}},\ }\href@noop {} {\bibfield  {journal} {\bibinfo  {journal}
  {Theoretical Economics}\ }\textbf {\bibinfo {volume} {8}},\ \bibinfo {pages}
  {1} (\bibinfo {year} {2013})}\BibitemShut {NoStop}%
\bibitem [{\citenamefont {Watts}(2002)}]{Watts2002}%
  \BibitemOpen
  \bibfield  {author} {\bibinfo {author} {\bibfnamefont {D.~J.}\ \bibnamefont
  {Watts}},\ }\href@noop {} {\bibfield  {journal} {\bibinfo  {journal}
  {Proceedings of the National Academy of Sciences}\ }\textbf {\bibinfo
  {volume} {99}},\ \bibinfo {pages} {5766} (\bibinfo {year}
  {2002})}\BibitemShut {NoStop}%
\bibitem [{\citenamefont {Caginalp}\ and\ \citenamefont
  {Doiron}(2017)}]{Caginalp2017}%
  \BibitemOpen
  \bibfield  {author} {\bibinfo {author} {\bibfnamefont {R.~J.}\ \bibnamefont
  {Caginalp}}\ and\ \bibinfo {author} {\bibfnamefont {B.}~\bibnamefont
  {Doiron}},\ }\href@noop {} {\bibfield  {journal} {\bibinfo  {journal} {SIAM
  Journal on Applied Dynamical Systems}\ }\textbf {\bibinfo {volume} {16}},\
  \bibinfo {pages} {1543} (\bibinfo {year} {2017})}\BibitemShut {NoStop}%
\bibitem [{\citenamefont {Geisler}(2003)}]{geisler03}%
  \BibitemOpen
  \bibfield  {author} {\bibinfo {author} {\bibfnamefont {W.~S.}\ \bibnamefont
  {Geisler}},\ }\href@noop {} {\bibfield  {journal} {\bibinfo  {journal} {The
  visual neurosciences}\ }\textbf {\bibinfo {volume} {10}},\ \bibinfo {pages}
  {12} (\bibinfo {year} {2003})}\BibitemShut {NoStop}%
\bibitem [{\citenamefont {Moreno-Bote}(2010)}]{moreno10}%
  \BibitemOpen
  \bibfield  {author} {\bibinfo {author} {\bibfnamefont {R.}~\bibnamefont
  {Moreno-Bote}},\ }\href@noop {} {\bibfield  {journal} {\bibinfo  {journal}
  {Neural computation}\ }\textbf {\bibinfo {volume} {22}},\ \bibinfo {pages}
  {1786} (\bibinfo {year} {2010})}\BibitemShut {NoStop}%
\end{thebibliography}%
\end{document}